\documentclass[english,aps,superscriptaddress, twocolumn]{revtex4-1}
\usepackage[T1]{fontenc}
\usepackage[latin9]{inputenc}
\usepackage{amsmath}
\usepackage{amssymb}
\usepackage{graphicx}
\usepackage{esint,booktabs}
\usepackage{multirow}

\makeatletter

\@ifundefined{date}{}{\date{6 June 2021}}
\makeatother

\usepackage{babel}
\begin{document}
	
	\title{GeV muon beams with picometer-class emittance from electron-photon collisions}
	
	\author{C. Curatolo} 
	\affiliation{Università degli Studi di Padova, via Marzolo 8, 35131 Padova, Italy}
	\affiliation{INFN-Padova, via Marzolo 8, 35131 Padova, Italy}
	
	\author{L. Serafini} 
	\affiliation{INFN-Milan, via Celoria 16, 20133 Milano, Italy}
	
\begin{abstract}

One of the challenge of future muon colliders is the production of muon beams carrying high phase space densities. In particular the muon beam normalised transverse emittance is a relevant figure of merit to meet luminosity requests. A typical issue impacting the achieved transverse emittance in muon collider schemes so far considered is the phase space dilution caused by coulomb interaction of primary particles propagating into the target where muons are generated. In this study we present a new scheme for muon beam generation occurring in vacuum by interactions of electron and photon beams. Setting the center of mass energy at about twice the threshold (i.e. around $350$ MeV) the normalised emittance of the muon beam generated via muon pair production reaction ($e^-+\gamma \rightarrow e^-+\mu^+/\mu^-$) is largely independent on the emittance of the colliding electron beam and is set basically by the excess of transverse momentum in the muon pair creation. In absence of any other mechanism for emittance dilution, the resulting muon beam, with energy in the range of few tens of GeV, is characterised by an ultra-low normalised transverse rms emittance of a few nm rad, corresponding to a geometrical emittance below $10$ pm rad. This opens the way to a new muon collider paradigm based on muon sources conceived with primary colliding beams delivered by $100$ GeV-class energy recovery linacs interacting with hard-X ray free electron lasers. The challenge is to achieve the requested luminosity of the muon collider adopting a strategy of low muon fluxes/currents combined to ultra-low emittances, so to largely reduce also the levels of muon beam-induced background. 
\end{abstract}

\maketitle

\section{Introduction}
Muon beam generation for muon colliders has been traditionally conceived using hadronic interactions that go through the pion production channel with consequent decay into muons: the total cross section for these reactions is quite large, assuring a wealth of muon population when intense proton beams are impinging on targets made of proper material \cite{MAP}. Unfortunately the coulomb interaction of primary and secondary beams into target largely disrupts the phase space of the generated muon beam, diluting its emittance well above the upper limits set by collider luminosity requests, so that a challenging muon cooling process must be implemented via ionisation with the aim to restore the emittance levels down to values compatible for collider luminosity \cite{ref:mice}.
Alternatively, positron-electron annihilation into muon pairs has been proposed as a promising scheme to achieve directly at the muon source low emittance values \cite{LEMMA}. Investigations on this scheme are still in progress to assess its full potentiality: nevertheless the primary beam of $45$ GeV positrons has to interact with a solid target in order to achieve sufficient muon pair production rate, which in turns poses issues of target handling.
An exotic scheme of Hadron Photon Collider (HPC) was proposed to generate muon pairs through the reaction $p^-+\gamma \rightarrow p^-+\mu^+/\mu^-$: in this case the emittance of the muon beam was quite good, basically conserving the emittance of colliding proton and photon beams \cite{HPC, HPC2,HPC3}. The very small total cross section typical of muon photo-production reactions sets a severe limit to the muon flux achievable.
In this paper we revisit the basic concepts of that study, implementing the idea of muon photo-production in a different environment - that of electron-photon collisions. The reason why lays not really in a larger cross section of the $e, \, \gamma$ reaction with respect to $p, \,\gamma$ (the increase is quite modest), but in the possibility to run an electron-photon collider at much larger luminosity than a proton-photon collider as in the HPC scheme. As a matter of fact the focal spot size of a $100$ GeV-class electron beam colliding in-vacuum with a counterpropagating photon beam can be much smaller than that of a TeV proton beam stored in a machine like LHC, as well as that of an electron-positron collider final focus. The former is actually limited by the minimum beta achievable in a proton storage ring due to cm-long proton bunches, the latter is actually limited by the beam-beam interaction mutuated by the presence of colliding charged particles focused into a tight focal region (the so called beamstrahlung effect). In case of electron-photon collisions none of these effect are actually present - electrons and photons interact just through 2-particle QED reactions without any significant collective e.m. field effect impacting the beam propagation through the interaction region. So the maximum focusing achievable in the collision region is just dictated by final focus beam optics, electron and photon bunch lengths and transverse emittances of the two colliding beams.
In order to reach the Center of Mass (CM) available reaction energies in excess of $300$ MeV (where the total cross section exceeds few hundreds of nano barns) we are forced to consider electrons with energies higher than $100$ GeV and photons with energy in the $100-300$ keV range. Following the analysis pursued in HPC studies, in order to maximise the luminosity of an electron-photon collider in this range of particle-photon energies we are naturally addressed to consider a hard X-ray Free Electron Laser (FEL) as the brightest radiation source and an Energy Recovery Linac (ERL) to obtain the maximum electron beam current (A-class) combined with high brightness - namely short electron bunches (fs-class) with very small transverse normalised emittances (below $1$ mm mrad).
Since no concern is raised by beam-beam interaction effects, maximum luminosity is achieved running round beams at the collision aiming at minimum beta values at the final focus. We will show in the following discussion that close to state of the art performances both of FELs and ERLs are indicating the possibility to achieve an outstanding luminosity in the $e^--\gamma$ collider, up to $10^{41}$ cm$^{-2}$s$^{-1}$, which allows to generate up to $10^{11}$ muon pairs per second with geometrical emittances down to 5 pm rad.
The paper is organised as follows: in Section \ref{sec:theory} we summarize the theoretical aspects regarding the electron-photon beams collision and the main formulas we will use to analyse the various cases reported in Section \ref{sec:par}. In the latter we discuss the performances requests to primary beams in different scenarios of $e^--\gamma$ colliders, either using parasitically or semi-parasitically the layout of other projects under proposal in the High Energy Physics (HEP) future scenarios (CLIC \cite{CLIC}, FCC \cite{FCC}), or conceiving a dedicated scenario of colliding ERLs  (Section \ref{sec:el}). In Section \ref{sec:fel} we discuss how to achieve the requested FEL performances, running dedicated ERLs to drive the hard X-ray FEL, while in Section \ref{sec:pow} the power consumption demands are summarized. In Section \ref{sec:sim} we describe the results of Monte Carlo dedicated simulations to assess the phase space distributions and quality of the generated muon beams, its flexibility and dynamic range. Conclusions roughly depict the way towards a TeV muon collider based on $e^--\gamma$ collider muon source.

\section{Theory}\label{sec:theory}
We analyse here the most important aspects concerning the electron-photon beams collision at center of mass energies around 250-450 MeV. \\
The center of mass energy is given by
\begin{equation}
E_{CM}=\sqrt{2E_{e}h\nu -2(\underline{p}_{e}\cdot \underline{k}) +M_{e}^{2}}
\end{equation}
where $E_{e}, \, \underline{p}_{e}$ and $h\nu, \, \underline{k}$ are electron and photon energies and momenta respectively and $M_{e}=0.511$ MeV/c$^2$ is the electron mass (natural units are used i.e. c=1). Assuming a head-on collision within the electron and the incident photon with energy $h\nu<<E_{e}$ and supposing $M_{e}<<E_{e}h\nu$, the Lorentz factor of the center of mass is
\begin{equation}
\gamma_{CM}=\frac{E_{tot}^{LAB}}{E_{CM}}\simeq\frac{E_{e}+h\nu}{\sqrt{4E_{e}h\nu+M_{e}^{2}}}\simeq\frac{E_{e}}{2\sqrt{E_{e}h\nu}}= \frac{\sqrt{E_{e}}}{2\sqrt{h\nu}}
\end{equation} 
The CM threshold energy for the Muon Pair Production (MPP) is given by
\begin{equation}
E_{CM}^{th}=M_{e}+2M_{\mu}
\end{equation}
If for example $E_{e}=200$ GeV, the equality $E_{CM}^{th}=M_{e}+2M_{\mu}=\sqrt{4E_{e}h\nu+M_{e}^{2}}$
implies that $h\nu^{th}=56.6$ keV. The photon energy in the electron rest frame is 
$h\nu^{th'}=2\gamma_{e}h\nu^{th}=44.3$ GeV while the photon energy in the CM frame is $h\nu^{th*}=2\gamma_{CM}h\nu^{th} =106.5$ MeV.\\
Beside MPP, the other predominant reactions in the mentioned CM energy range are Triplet Pair Production (TPP - $e^-+\gamma \rightarrow e^-+e^+/e^-$) and Inverse Compton Scattering (ICS - $e^-+\gamma \rightarrow e^-+\gamma$). 
\begin{figure}[htbp]\centering
	\includegraphics [scale=0.5] {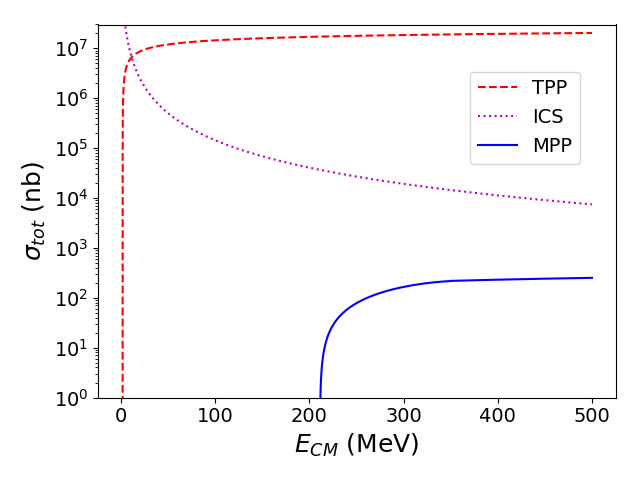}
	\caption{Total cross sections of possible reactions in electron-photon collision as function of $E_{cm}$ (MeV).}\label{tcs}
\end{figure}
Their total cross sections $\sigma_{tot}$ as a function of $E_{CM}$ are depicted in Fig.\ref{tcs}. 
The TPP cross sections calculated as in Ref. \cite{motz} is 
$$\sigma^{TPP}_{tot}=\sigma^{TPP}_{Bethe-Heitler}-\sigma^{TPP}_{Borsellino}$$
\begin{equation}\begin{split}
\begin{cases}
\sigma^{TPP}_{B-H}=\alpha r_{e}^2&(3.11 \, \ln(2k_{e})-8.07)\\
\sigma^{TPP}_{Bors}=\frac{\alpha r_{e}^2}{k_{e}}&\left(\frac{4}{3} \, \ln(2k_{e})^3-3 \, \ln(2k_{e})^2+\right.\\
&\left.\mbox{ }+ 6.84 \, \ln(2k_{e})-21.51\right)
\end{cases}
\end{split}
\end{equation}
where $\alpha=1/137$ is the fine-structure constant, $r_e^2=0.079$ barn is the classical electron radius squared and $k_{e}=2 \gamma_e h\nu/M_e$.\\
In a similar way, the MPP cross section can be obtained from the one above by substituting $r_e$ with $r_{\mu}= r_{e}(M_{e}/M_{\mu})$ and $k_e$ with $k_{\mu}=2 \gamma_e h\nu/M_{\mu}$. Its value is well-defined if $E_{CM}^2/M_{\mu}^2>9$ and we also find that $E_{CM}=\sqrt{2k_{\mu}M_{\mu}M_{e}}$.\\
The ICS cross section, as function of the electron recoil $X=2 h\nu'/M_e=4\gamma_e h\nu/M_e$, is given by \cite{ics}
\begin{equation}\begin{split}
\sigma&^{ICS}_{tot}=2\pi r_e^2\frac{1}{X} \cdot\\
&\cdot \left[\left(1-\frac{4}{X}-\frac{8}{X^2}\right)\ln(1+X)+\frac{1}{2}+\frac{8}{X}-\frac{1}{2(1+X)^2}\right].\label{stot}
\end{split}
\end{equation}
The luminosity of the $e^--\gamma$ collider is defined as
\begin{equation}
\mathcal{L}= \frac{N_{e}\,N_{ph}\,r}{2\,\pi \, \sqrt{\sigma_{x_e}^2+\sigma_{y_e}^2}\sqrt{\sigma_{x_{ph}^2}+\sigma_{y_{ph}}^2}} \,[\mbox{cm}^{-2} \mbox{ s}^{-1}]
\end{equation}
being $N_{e}, N_{ph}$ the number of electron and photons per bunch, $r$ the repetition rate of the collisions and $\sigma_{x_e}, \sigma_{y_e}, \sigma_{x_{ph}}, \sigma_{y_{ph}}$ the transverse dimensions of the electron and the photon beams respectively. Therefore the number of muon and electron pairs and ICS events per second are
\begin{equation}\begin{split}
\mathcal{N}_{\mu^{\pm}}= \mathcal{L}\cdot \sigma^{MPP}_{tot}(E_{CM}) \,[\mbox{s}^{-1}]\\
\mathcal{N}_{e^{\pm}}= \mathcal{L}\cdot \sigma^{TPP}_{tot}(E_{CM}) \,[\mbox{s}^{-1}]\\
\mathcal{N}_{ICS}= \mathcal{L}\cdot \sigma^{ICS}_{tot}(E_{CM}) \,[\mbox{s}^{-1}]
\end{split}
\end{equation}

\noindent The transverse normalized emittance of the produced muons is determined by the intrinsic thermal contribution of the reaction and by the features of the incoming electron beam. It can be described by
\begin{equation}
	\epsilon_{\mu}^n\simeq\frac{2}{3}\sigma_{0}\frac{\sqrt{E_e \, h\nu}-M_{\mu}}{M_{\mu}}+\frac{\langle\gamma_{\mu}\rangle \epsilon_{e}^n}{\gamma_e}
	\label{em}
\end{equation}
where $\epsilon_{e}^n$ is the incoming electron beam transverse normalized emittance, $\langle\gamma_{\mu}\rangle$ the mean energy of the muon beam, $\sigma_0=\sqrt{\left(\sqrt{\sigma_{x_e}^2+\sigma_{y_e}^2}\sqrt{\sigma_{x_{ph}}^2+\sigma_{y_{ph}}^2}\right)/2}$ and the first addendum represents the normalized thermal emittance of the muon beam. In the cases we will consider, the emittance of the incoming electron beam has basically no impact on the muon beam one: the second part of Eq. \ref{em} is negligible because $\langle\gamma_{\mu}\rangle \ll \gamma_e$.

\section{$e^--\gamma$ collider scenarios}\label{sec:par}
We have conceived a few scenarios for $e^--\gamma$ colliders achieving ultra high luminosities, considering various technologies for the electron accelerators involved in the generation of the primary electron beam and in the generation of the FEL hard X-ray photon beam. An overarching goal is the quest of very high sustainability: such levels of luminosities typically require beam power levels that are hardly compatible with reasonable impacts on society and governmental approvals when power consumption of the whole scenario start to exceed the so called GW scale, that should be considered as an unsurpassable borderline.
The resulting criterion for the choice of the electron accelerators is the following: since the generation of muon pairs in the collision between the primary electrons and the FEL photons is perturbative on the primary beams, with negligible losses of beam particles and very small energy loss in the beams, a strategy for beam power recirculation or recovery is mandatory. The main source of beam power loss in these scenarios is set by the amount of synchrotron radiation power generated at any electron beam deflection or recirculation. Four scenarios have been analysed and described in detail in Table \ref{tab}: two of them are based on parasitic use of machines that are already in a proposal stage in the HEP future strategy, i.e. CLIC and FCC-ee. Two further scenarios are based on PERLE-like ERLs \cite{perle1,perle2}): one, named FCC-twin Linac, is integrated in FCC tunnels system, the other, named Twin Linac, is based on an ILC schematics \cite{ILC,shil}. Only these scenarios satisfy the luminosity requirements but we report for sake of comparison and point of reference also the performances of parasitic operations of CLIC and FCC-ee.

\begin{figure}[ht]\centering
	\includegraphics [scale=.6] {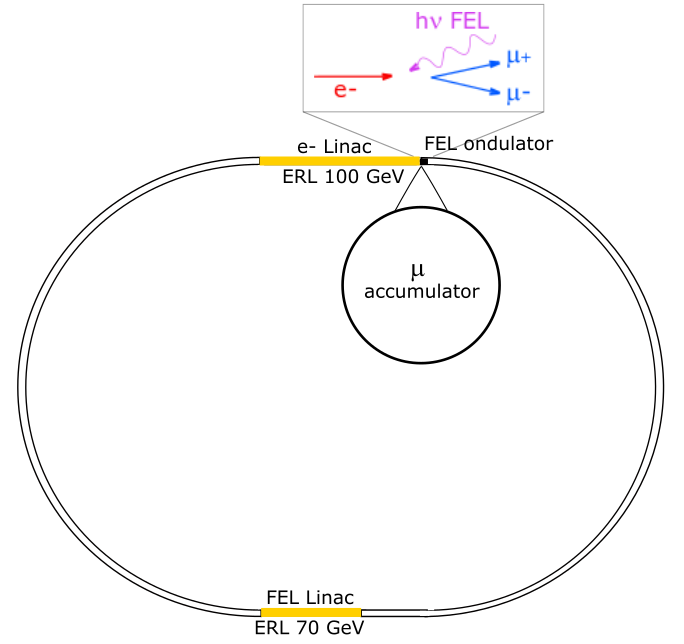}
	\caption{FCC-twin Linac scheme: FCC ring with ERL insertions. Primary $e^-$ accelerated up to $100$ GeV collide with the counterpropagating FEL producing $\mu^\pm$. Both $e^-$ beams (primary and FEL) energies are recovered. $\mu^\pm$ accumulated and accelerated in a separate ring.}\label{fcctwin}
\end{figure}
\begin{figure*}[ht]\centering
	\includegraphics [scale=.5] {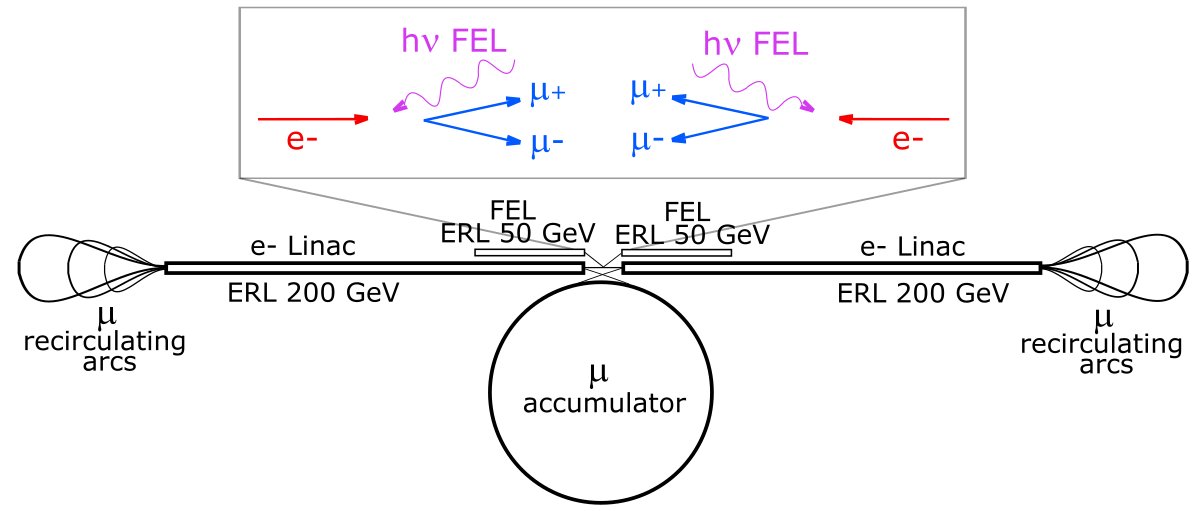}
	\caption{Twin Linac scheme. Primary $e^-$ accelerated up to $200$ GeV collide with the counterpropagating FEL producing $\mu^\pm$. Both $e^-$ (primary and FEL) are decellerated in the opposite linacs and the energy recovered. A selected fraction of $\mu^\pm$ are injected in the opposite linac and accelerated before stogare and collision in a ring.}\label{twin}
\end{figure*}
\begin{table*}[ht]	
	\caption{Parameters table of the four schemes: FCC-twin Linac, CLIC parasitic, FCC-ee parasitic, Twin Linac. Primary beams features and peak luminosities are reported.}\label{tab}
	\begin{tabular}{lcccc} 
		\toprule
		\midrule
		& FCC-twin Linac \hspace{.1cm} & CLIC parasitic \hspace{.1cm} & FCC-ee parasitic \hspace{.1cm} & Twin Linac \\
			\midrule
		Energy  $e^-$ beam (GeV)    & $100$   & $1500$     & $200$    & 200    \\
		Bunch charge (pC)      & 250    & 580  &  43200     & 250    \\
		Electrons per bunch    & 1.6 $10^9$ & 3.7 $10^9$& 2.7 $10^{11}$ & 1.6 $10^9$ \\
		Repetition rate (MHz)  & 800  (CW)  & 0.005    & 0.1    & 800  (CW)  \\
		Average Current (mA)   & 200    & 0.003    & 5.4    & 2$\times$200  \\
		Nominal beam power (GW)    & 20     & 0.0045   & n.a.   & 2$\times$40  \\
		Energy loss 1 turn @ 15 km r (MeV)     & 593    & n.a.     & 9500   & n.a.   \\
		Synch. rad. 1 turn @ 15 km r (MW)   & 119    & n.a.     & 51     & n.a.   \\
		Beam power recovery fraction   & 99.9\%    & n.a  & n.a.   & 99.9\%    \\
		Beam power loss (MW)   & 20     & 4.5  & n.a.    & 2$\times$40   \\
		Bunch length (psec)    & 0.3    & 0.15     & 12     & 0.3    \\
		$\epsilon^n_x$ (m rad)& 4 $10^{-7}$ & 6.6 $10^{-7}$ &520 $10^{-6}$&4 $10^{-7}$\\
		$\epsilon^n_y$(m rad)  & 4 $10^{-7}$ & 3 $10^{-8}$   & $10^{-6}$ & 4 $10^{-7}$ \\
		$\beta_x$ (mm)    & 0.2    & 0.6  & 1.6    & 0.2    \\
		$\beta_y$ (mm)    & 0.2    & 12   & 1.6    & 0.2    \\
		$\sigma_x$ (nm)    & 20     & 10   & 1442   & 14     \\
		$\sigma_y$ (nm)    & 20     & 10   & 63     & 14     \\
			\midrule
		FEL photon energy (keV)    & 300    & 30   & 75     & 150    \\
		Photons per pulse  & $10^{12}$   & 1.6 $10^{14}$ & 2 $10^{13}$ & 5 $10^{12}$ \\
		$\epsilon_{x,y}$ (m rad) & 3 $10^{-13}$& 3 $10^{-12}$  & 1.2 $10^{-12}$& 6 $10^{-13}$ \\
		Focal spot size (nm)  & 20     & 10   &  (1400 $\times$ 60)        & 14  \\
		Repetition rate (MHz)  & 800    & 0.005    & 0.1    & 800    \\
		FEL beam power (MW)    & 40     & 0.00375  & 0.025  & 2$\times$100  \\
		FEL $\rho$    & 3 $10^{-4}$    & $10^{-3}$  & 7 $10^{-4}$     & 5 $10^{-4}$ \\
		FEL efficiency (tapering)  & 0.5\%  & 5\%  & 2\%    & 1\%    \\
		FEL  $e^-$ beam av. curr. (mA)  & 200    & 0.0025   & 0.025  & 200    \\
		FEL  $e^-$ beam bunch ch. (pC)     & 250    & 500  & 250    & 250    \\
		FEL  $e^-$ beam energy (GeV)   & 70     & 30   & 50     & 50     \\
		FEL  $e^-$ beam power (MW)  & 14000     & 0.075 & 1.25    & 2$\times$10000   \\
		Beam power recovery fraction   & 99.9\%     & n.a.     & n.a.   & 99.9\%  \\
		Beam power loss (MW)   & 14     & 0.075    & 1.25   & 2$\times$10   \\
			\midrule
		Total beam power loss (MW) & 202    & n.a.     & n.a.   & 200    \\
		Peak Luminosity $e^--\gamma$ (cm$^{-2}$s$^{-1}$) & 2.5 $10^{40}$ & 2 $10^{38}$ & 3 $10^{35}$ & 2$\times$2.5 $10^{41}$\\
		\midrule 
		\bottomrule
	\end{tabular}
\end{table*}
The final goal of generating a suitable beam of muons to be used in a TeV-scale muon collider is accomplished by the Twin Linac scenario based on a twin $200$ GeV ERL system coupled to a twin $50$ GeV FEL ERL system, with residual beam power loss (after recovery, and taking into account all losses of electron beam and FEL photon beam) of about $200$ MW, together with an outstanding $4.5$ nm rad normalised emittance of the muon beam at $10^{11}$ muon pairs per second.

\subsection{The primary electron beam}\label{sec:el}
We illustrate here the rationale of the first and fourth column listed in Table \ref{tab} for the primary electron beam parameter case. As a matter of fact the intermediate cases, based on CLIC and FCC-ee, are just taken from electron beam parameter sets reported in the literature and they are listed just to show how far they are from the target value for luminosity.
Instead, the first case, based on a ERL integrated in the FCC tunnel (with proper tunnel reshaping to achieve a race-track layout), and the fourth case, based
on a dedicated twin Linac layout, are both capable to meet the tight demands on luminosity.

An important consideration concerns the peak properties fo the primary electron beam for both these FCC-tL and TL cases: the parameters listed in the table
are quite consistent with present state of the art for electron beams, with a rms normalised transverse emittance ($0.4$ mm mrad, round beam)  quite consistent
with the accelerated bunch charge ($250$ pC). Also the value chosen for the beta function at the collision point ($0.2$ mm) is very close to state of the art performances. That allows to match the spot size of the FEL photon beam down to $14$ nanometers.

The power budget for the primary electron beams is as follows: in case FCC-tL the power loss is dominated by the synchrotron emission along the two recirculating arcs
(basically one turn of the FCC ring with about $15$ km curvature radius), summing up to $119$ MW. In addition, considering a recovery efficiency of $99.9\%$ in the ERL, further $20$ MW of electron beam power is lost, bringing to a total beam power loss of about $139$ MW.
In the Twin Linac scenario no significant synchrotron radiation is expected, therefore the primary beam power loss is set by the ERL efficiency, bringing up to $2 \times 40 = 80$ MW.

\subsection{High efficiency hard X-ray FELs}\label{sec:fel}

The photon beam needed to achieve ultra high luminosity in the $e^--\gamma$ collider is unique: it must carry an outstanding number of photons per pulse at the same repetition rate of the primary electron beam. It must also match the ultra tight focus spot size at collision of the primary electron beam, set by its very short beta function value at the focus (hundreds of microns), in the range of a few tens of nanometers. The only radiation source capable to meet these demanding requirements is a FEL driven by a dedicated ERL, and operated in SASE mode with tapering, as illustrated in Ref. \cite{emma}. Efficiency in the range of a few percent is achievable, bringing to a number of photons per pulse as listed in Table \ref{tab}, according to photon energy. The partial coherence of the amplified FEL radiation makes also possible to focus its photon beam down to nanometer spot sizes, as discussed in Ref. \cite{matsu}. This is the second crucial property of FELs that makes possible to foresee a luminosity for the $e^--\gamma$ collider best scenario (based on a system of two twin ERLs at $200$ GeV electron energy, combined to two twin $50$ GeV ERLs driving the FELs) of up to $2 \times 2.5 \,10^{41}$ cm$^{-2}$s$^{-1}$.

The FEL beams considered in this study carry an impressive amount of photon beam power: running at $800$ MHz in CW mode, the photon number per second exceeds $10^{21}$. With photon energies in the range of tens to hundreds keV that means up to $100$ MW radiation beam power. Since the FEL efficiency, considering the special mode of FEL operation as illustrated in Ref. \cite{emma}, is about $1\%$, the power carried by the electron beam driving the FEL must be of the order of $100$ MW/$0.01 = 10$ GW, as listed in Table \ref{tab}. Therefore, the power budget for the Linac driving the FEL is as follow: in case FCC-tL we have a synchrotron radiation accounting for $29$ MW, a recovery beam efficiency for $14$ MW and a FEL beam power loss of $40$ MW. Assuming that $50\%$ of the radiation power may be recovered back as AC power, the total power budget is $29+14+20 = 63$ MW of beam power loss. In case of Twin Linac we have twice the recovery beam power loss of $10$ MW, and twice $50\%$ of $100$ MW, summing up to $120$ MW.

\subsection{Power budget}\label{sec:pow}
Summing up the power budget of primary electron beam to that of FEL driving electron beam we get a total beam power loss of $202$ MW for the FCC-tL case, and $200$ MW for the Twin Linac case. An expected efficiency beam-to-plug not smaller than $20\%$, actually in the range $20-40\%$, would set the AC power bill in the $250$ MW - $1$ GW range.

One should note that the power transferred from the primary colliding electron/photon beams into the secondary beams of muon pairs, photons, and electron-positron pairs is quite negligible compared to the stored power into the colliding beams at the collision point.
As a matter of fact, in case of Twin Linac scenario, the muon pair beams ($10^{11}$ per second at an average energy of 50 GeV) are taking out only $1$ kW of beam power. The power taken by the back-scattered Compton gammas ($8 \,10^{12}$ per second at an average energy of $200$ GeV) is $250$ kW, and the power taken by the electron-positron pair produced ($10^{16}$ per second at an average energy of $1$ GeV) is about $1.6$ MW. All these are quite negligible with respect to the total power loss quoted in Table \ref{tab} ($200$ MW), that is dominated by ERL efficiency and FEL photons (and synchrotron emission in case of FCC-tL).
Furthermore, in case of Twin Linac scenario, the real estate on top of the underground whole Linac bunker, that extends over about $6$ km$^2$ (nearly $30$ km length times $200$ m wide), if covered by solar panels would make the whole system almost self-sustained by about $900$ MW AC power generated by sun light.

\section{Simulation results for muon beams}\label{sec:sim}
\begin{figure}[h]\centering
	\includegraphics [scale=0.55] {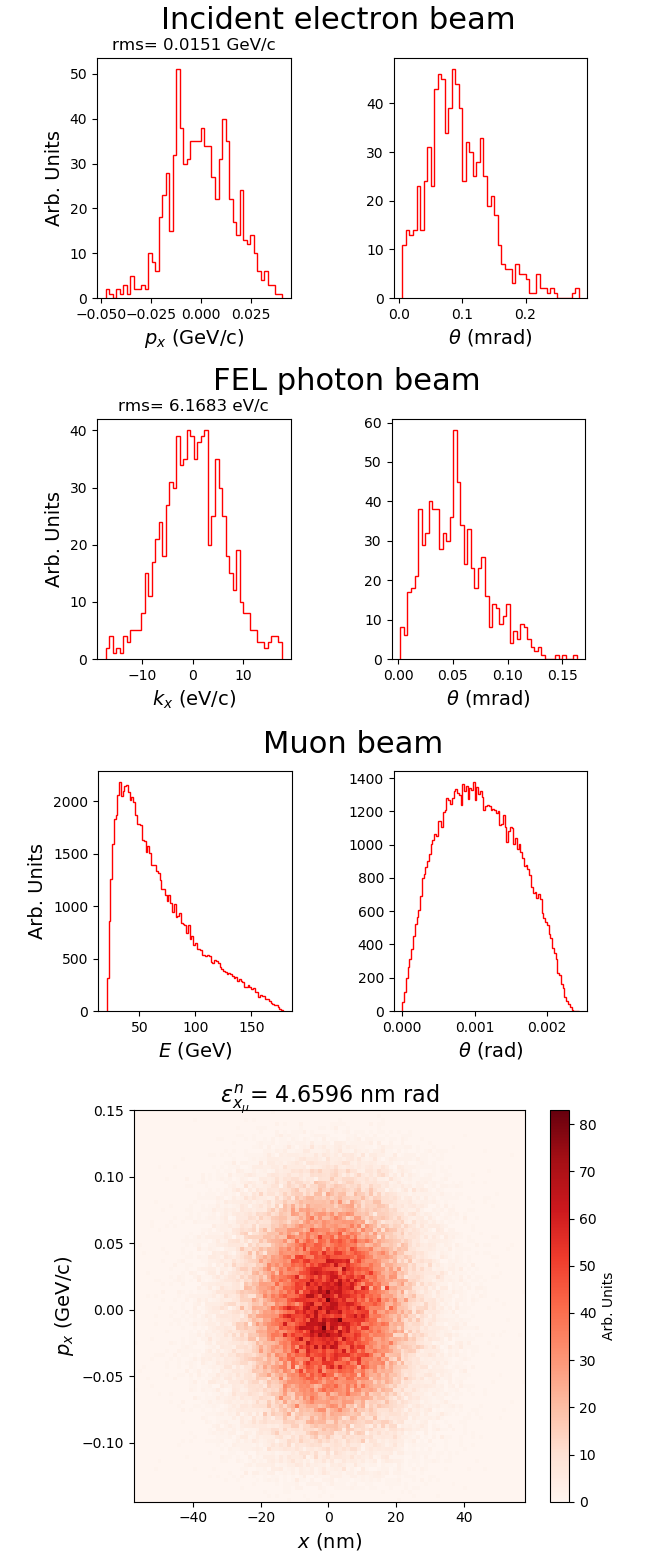}
	\caption{Twin Linac scheme: incident $e^-$ beam, FEL photon pulse and MPP features.}
	\label{TL1}
\end{figure}

\begin{figure}[h]\centering
	\includegraphics [scale=0.52] {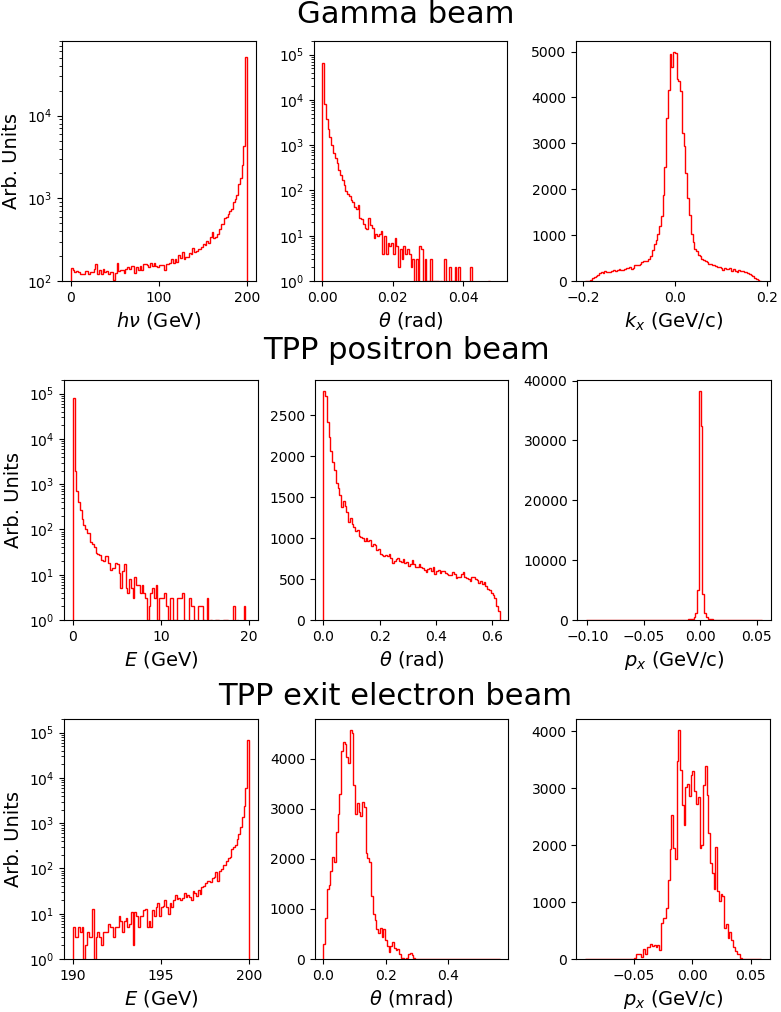}
	\caption{Twin Linac. Emitted ICS photon, TPP positron and initial electron beam after TPP reaction beam features.}
	\label{TL2}
\end{figure}

The significant values concerning the emitted particles for the four analysed schemes are reported in Table \ref{simres}. The best option is represented by the Twin Linac scheme that produces  $2 \times 5.4 \, 10^{10}$ muon pairs per second. 
\begin{table}[h]
	\caption{Emitted particles characteristics.  $^{*}$ TPP and MPP cross section values for the FCC-twin Linac and Twin Linac schemes are calculated as in Section \ref{sec:theory} while Whizard returns values lower by a factor $\sim 4$. $^{**}$ The MPP total cross section in the FCC-ee parasitic scheme is not well-defined since $E_{CM}^2/M_{\mu}^2=5.4$.\label{simres}}
	\begin{tabular}{lcccc} 
		\toprule
		\midrule
		& FCC-tL & CLIC par & FCC-ee par & TL \\
		\hline
		$E_{CM}$ (MeV)& $ 346$ & $424 $ & $244$ & $ 346$ \\
		\hline
		$\sigma^{TPP}_{tot}$ (mb)& $ 19^{*}$ & $19.7 $ & $17.7$ & $ 19^{*}$ \\
		\hline
		$\sigma^{ICS}_{tot}$ ($\mu$b)& $ 14.7 $ & $ 10$ & $ 27.9$ & $ 14.7 $ \\
		\hline
		$\sigma^{MPP}_{tot}$ (nb)& $ 216^{*}$ & $235$ & $183^{**}$ & $216^{*}$ \\
		\hline
		$\mathcal{N}_{e^{\pm}}$ (s$^{-1}$) & $4.8 \, 10^{14}$ & $2.9 \, 10^{11} $ & $4.2 \, 10^{10}$ & $2 \times 4.8 \, 10^{15}$ \\
		\hline
		$\mathcal{N}_{ICS}$ (s$^{-1}$)& $ 3.7 \, 10^{11}$ & $1.5 \, 10^{8}$ & $6.6 \, 10^7$ & $2 \times 3.7 \, 10^{12}$ \\
		\hline
		$\mathcal{N}_{\mu^{\pm}}$ (s$^{-1}$) & $5.4 \, 10^9 $ & $3.4 \, 10^6$ & $ 4.3 \, 10^5$ & $2 \times 5.4 \, 10^{10}$ \\
		\hline
		$\epsilon^n_{x_{\mu}}$ (nm rad) & $6.7$ & $4.1$ & $873.9$ & $4.6$ \\
		\hline
		$\mathcal{N}_{\mu^{\pm}}/\epsilon^n_{x_{\mu}}$&\multirow{ 2}{*}{ $806$} &\multirow{ 2}{*}{ $0.8$} &\multirow{ 2}{*}{ $0.0005$} &\multirow{ 2}{*}{ $2 \times 11739$} \\ 
		($10^{15}$ m$^{-1}$s$^{-1}$)&  &  &  & \\ 
		\midrule
		\bottomrule
	\end{tabular}
\end{table}
\begin{figure}[h]\centering
	\includegraphics [scale=0.52] {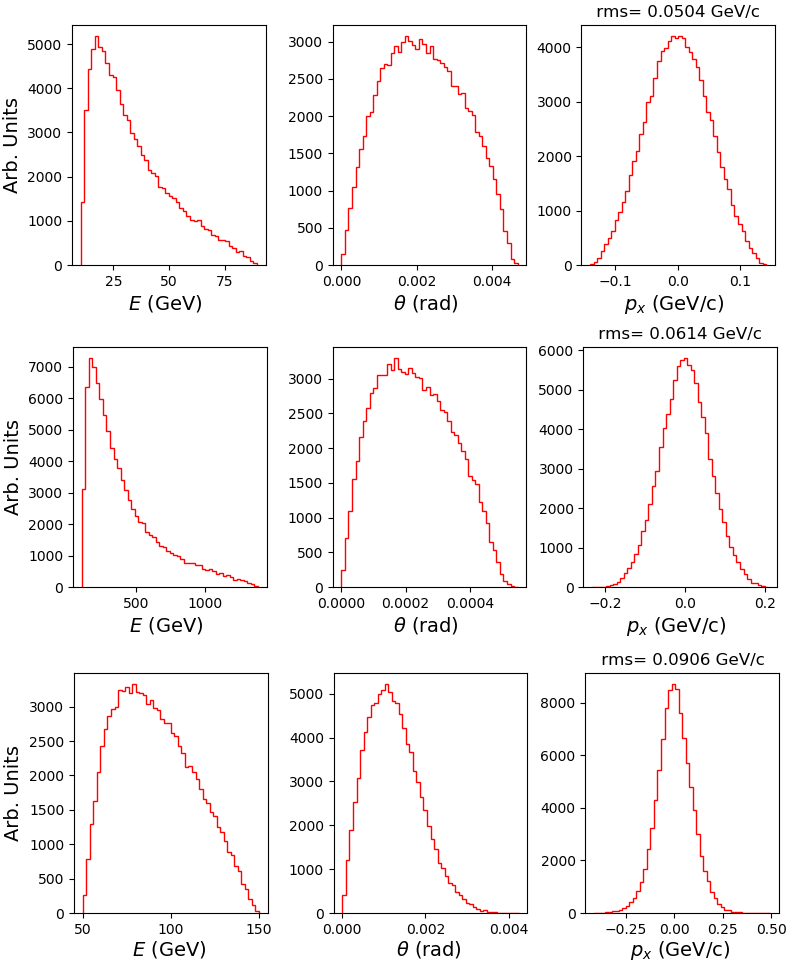}
	\caption{MPP features. Top line: FCC-Twin Linac scheme; middle line: CLIC parasitic scheme; bottom line: FCC-ee parasitic scheme.}
	\label{mu1}
\end{figure}
The MPP has been simulated by means of the Whizard event generator \cite{whizard, whizard1}, run in such a way to take into account the incoming beams features. The muon beam transverse normalized emittance for all the schemes is also reported in Table \ref{simres} and the number of muons per second divided by the emittance, crucial figure of merit as suggested in \cite{frank}, has been calculated. The best calculated value for muon beam normalized transverse emittance is $4.6$ nm rad: this value compares to the analytical prediction of Eq. \ref{em} giving $5.9$ nm rad.
The emitted muon beams features for the FCC-twin Linac, the CLIC parasitic and the FCC-ee parasitic schemes are reported in Fig. \ref{mu1}. The energy, angle and transverse momentum distrubutions are shown.
The Twin Linac option has been analysed in detail. The incoming electron beam, the FEL photon pulse and the MPP features are displyed in Fig. \ref{TL1}. 
The outstanding value of the transverse normalized emittance combined with the number of muon pair per second returns the value $\mathcal{N}_{\mu^{\pm}}/\epsilon^n_{x_{\mu}}=2 \times 1.17 \,10^{19}$ m$^{-1}$s$^{-1}$. If we consider the muons around the energy distribution peak of $50$ GeV corresponding to a $10\%$ rms relative energy spread, the $20\%$ of the produced muons are selected (with a longituadinal emittance value of $\sim4.5$ mm). The above mentioned coefficient corresponding to this selection is $\mathcal{N}_{\mu^{\pm}}/\epsilon^n_{x_{\mu}}=2 \times 2.34 \,10^{18}$ m$^{-1}$s$^{-1}$, comparable with the best option of the Gamma Factory \cite{GF1,GF2} combined with LEMMA \cite{LEMMA} analysed in Ref. \cite{frank}. 
Top line of Fig. \ref{TL2} shows the energy spectrum of the ICS photons produced in the collision: as expected in a high electron recoil regime, the energy is peaked around $200$ GeV. The TPP, which is the most probable collateral reaction, would in this case involve  $2 \times 4.8 \, 10^{15}$ primary electrons, still a small fraction of the total ($2 \times 10^{18}$ per second). Moreover, the bottom line of Fig. \ref{TL2} shows that most of the primary electrons involved in TTP would face a negligible energy loss since the $e^+/e^-$ pairs are mainly generated at very low energy (see middle line).
 
A useful characteristic of the Twin Linac layout shown in Fig. \ref{twin} is the possibility to accelerate muons in the same Linacs used to accelerate the two twin primary electron beams. This is possible thanks to the very small emittance of muons and the non-interacting muon-electron beams. Beam optics could rely on RF focusing that would be effective both on muons and on electrons, as further analysed in a future work. Linac acceleration of muons would allow to bring them up rapidly to the TeV kinetic energy range requested by muon collider physics, just in a few passes (each Twin Linac pass is $400$ GeV energy gain) through the Linac twin system, using proper muon recirculation arcs.

\section{Conclusions}
We described a muon source based on electron-photon collisions at ultra-high luminosity, capable to reach muon fluxes up to a few $10^{11}$ muon pairs per second at an outstanding normalized transverse emittance of a few nm rad, with muon beam energies peaked at $50$ GeV. The electron-photon collider is based on a primary electron Linac with energy in the $100-200$ GeV range and a FEL Linac driver with FEL photon energy in the $100-300$ keV range. Extremely large beam power (both electron and photon beams) are requested to achieve the ultra-high luminosity needed, in excess of $10^{40}$ cm$^{-2}$s$^{-1}$. Since the electron-photon collisions transfer only a very small amount of power from the primary beams into the secondary beams, an efficient energy recovery must be implemented in the scenario, so to reduce the amount of beam power loss down to hundreds MW level, from $100$ GW beam power stored in primary beams at collision. This is the main challenge of such a muon source, together with challenging beam collision spot size, in the few ten nanometers range, and handling of an extremely large FEL photon beam power.
Two scenarios have been analysed in details: one adopting the future FCC tunnel, with foreseen $15$ km radius of curvature, so to host in a slightly modified race-track geometry the two Linacs generating the primary electron beam and the FEL driving electron beam. This scenario is limited in performances by the huge amount of power lost due to synchrotron radiation emission in the arcs. The second scenario is based on a twin array of Linacs arranged face-to-face, providing  both the primary electron beam and the FEL driving beam: the performances of this scenario are well much higher in terms of muon flux, for the same beam power loss. Nevertheless, in this case it is of paramount importance to find a strategy of beam energy recovery with counterpropagating electron beams in the same Linac. This is the original ERL layout conceived by M. Tigner long time ago (see Ref. \cite{tigner}), subject to R\&D on twin super conducting cavities accomodating counterpropagating beams, but not yet demonstrated \cite{marix, uh}.
Further studies on the feasibility of these scenarios are necessary to assess the achievable luminosity of a muon collider based on this muon source, depending on the kind of accumulator scheme to combine with such muon source scenario. The promise is to achieve the requested luminosity using much lower muon beam currents with respect to other schemes, thanks to the very low emittance, which would alleviate significantly the issue of muon beam-induced background. Additional potentialities are also to be studied and carefully analysed, e.g. the simultaneous acceleration of muon beams and primary electron beams in the main Linacs, possible polarization of muon beams if polarized primary electron beams are used (FEL photons are naturaly polarized), parasitic use of intense GeV-class positron beams (up to $10^{16}$ pairs per second) generated in the electron-photon collisions, and of $100$ GeV-class intense (in excess of $10^{12}$ per second) monochromatic photons.

\end{document}